\documentstyle[12pt,epsf]{article}
\advance\topmargin by -3\baselineskip
\advance\textheight by 4\baselineskip
\begin{document}
\newcommand{\be}{\begin{equation}}
\newcommand{\ee}{\end{equation}}
\newcommand{\bea}{\begin{eqnarray}}
\newcommand{\eea}{\end{eqnarray}}
\newcommand{\ba}{\begin{array}}
\newcommand{\ea}{\end{array}}
\newcommand{\al}{\alpha}
\newcommand{\lb}{\label}
\newcommand{\ct}{\cite}
\newcommand{\ra}{\rightarrow}
\newcommand{\lab}{\Lambda_b}
\newcommand{\la}{\Lambda}
\newcommand{\dbn}{\bar{D}^o}
\newcommand{\de}{\delta}
\begin{center}
{\bf $\Lambda_b \rightarrow \Lambda + D (\bar{D}^0)$ decays and CP-violation} \\
Fayyazuddin\footnote{E-mail: fayyaz@hep-qau.sdnpk.undp.org} \\
Physics Department, Quaid-i-Azam University, Islamabad, Pakistan \\
\end{center}

\begin{abstract}
It is shown that interference of the amplitudes for the decays 
$\Lambda_b \rightarrow \Lambda D$ 
and 
$\Lambda_b \rightarrow \Lambda \bar{D}^0$ 
gives rise to $CP$-violation.
\end{abstract}
The decays $\Lambda_b \rightarrow \Lambda D^o$ and 
$\Lambda_b \rightarrow \Lambda \bar{D}^o$ are interesting for the
following reasons. (i) The Hamiltonian for the decay
$\Lambda_b \rightarrow \Lambda \bar{D}^o$ involves the weak
phase $ \hat{\gamma}$. This has direct implication for CP-Violation.
(ii) The factorization contribution for these decays is suppressed
due to the colour factor $C_2 + \xi C_1$ and the form factors
$g_V(m_{D^2})$ and $g_A(m_{D^2})$. In reference [1], we have recently 
shown that $g_V(m_{D^2})= g_A(m_{D^2}) = f_1(m_{D^2})= -0.120$ and 
that the contribution of second form factor is negligible to the decay 
widths and asymmetry parameter $\alpha$ (In fact $f_2/f_1 =0.129$).
(iii) The baryon $\prime{\Xi^o}_c$ and $\Xi_c^o$ poles contribute
only to the decay  $\Lambda_b \rightarrow \Lambda D^o$. The pole 
contribution can be evaluated in the $W$-exchange model [2].

The pole contributes to the p-wave amplitude and numerically its 
contribution is compareable to that of factorization. This has 
dramatic affect on the asymmetry parameter $\alpha$.
The asymmetry parameter $\alpha_{\bar{D}} = -1$, since only factorization
contributes to this decay where as $\alpha_D$ can be as low as zero
depending on the relative strength of the pole and factorization
contributions.

In the Wolfenstein parameterization [3] of CKM matrix [4], the weak
phases $\hat{\alpha}$, $\hat{\beta}$ and $\hat{\gamma}$ are given
by $\tan \hat{\alpha} = \frac{\eta}{\eta^2 - \rho(1-\rho)}$,
$\tan \hat{\beta} =\frac{\eta}{1-\rho}$ and
$\tan \hat{\gamma} = \frac{\eta}{\rho}$.
The weak phases have been denoted with $\hat{}$ to distinguish them from
asymmetry parameters $\alpha$, $\beta$ and $\gamma$. In the
two body non-leptonic decays of baryons, the angles $\phi$ and $\Delta$
defined as $\tan \phi = \beta/ \gamma$ and $\tan \Delta = - \beta / \alpha$
[5] give direct information about the $CP$ violation.
Due to interference of amplitudes for the decays $\Lambda D^o$ and
$\Lambda \bar{D}^o$, the decay amplitudes for  
$\Lambda_b \rightarrow \Lambda D_{\pm}^o $, where
$D_{\pm}^o = \frac{1}{\sqrt{2}} (D^o \pm \bar{D}^o)$
would make angles $\phi_{\pm}$ and $\Delta_{\pm}$ non zero in the absence
of final state interactions, if the pole contribution is not negligible.
These decays have also been discussed in reference [6], but the overlap
is minimal.

The decays $\Lambda_b \rightarrow \Lambda D^o$ and
$\Lambda_b \rightarrow \bar{D}^o$ are described by the effective
Hamiltonian [7]
 
\bea
H_{eff} (\Delta B = 1) &=& \frac{G_F}{\sqrt{2}} 
	\left[ V_{cb} V^{*}_{us} \left(C_1 O^c_1 + C_2 O_2^c \right) \right. \nonumber \\
	&+& \left. V_{ub} V^{*}_{cs} \left(C_1 O^u_1 + C_2 O_2^u \right) \right],   
\eea        
where $C_i$ are Wilson coefficients evaluated at the normalization scale $\mu$;  
the current-current operators $O_{1,2}$ are 
\bea
O^c_1 &=& \left(\bar{c}^{\alpha} b_{\alpha} \right)_{V-A} 
	\left(\bar{s}^{\beta} u_{\beta} \right)_{V-A} \nonumber \\
O^c_2 &=& \left(\bar{c}^{\alpha} b_{\beta} \right)_{V-A} 
	\left(\bar{s}^{\beta} u_{\alpha} \right)_{V-A} 
\eea
and $O^u_i$ are obtained through replacing $c$ by $u$. Here $\alpha$ and 
$\beta$ are $SU(3)$ color indices while 
$\left(\bar{c}^{\alpha} b_{\alpha} \right)_{V-A} =
	\bar{c}^{\alpha} \gamma_{\mu} \left(1 + \gamma_5 \right) b_{\beta}$  
etc. We take \ct{8}        
\bea
C_1 &=& 1.117 \nonumber \\
C_2 &=& - 0.257.
\eea 

The factorization contributions to the $s$-wave and $p$-wave amplitudes 
$A$ and $B$ are given by 
\bea
A_D     &=& - \, \frac{G_F}{\sqrt{2}} 
		V_{cb} V^{*}_{us} 
		\left(C_2 + \xi C_1 \right) 
		F_D \left(m_{\Lambda_b} - m_{\Lambda} \right) 
		g_V (m_D^2) \nonumber \\
	&=& \frac{G_F}{\sqrt{2}} (A \lambda^3) a_f \nonumber \\
B_D &=& \frac{G_F}{\sqrt{2}} (A \lambda^3) [b_f + b_p] \nonumber \\
A_{\bar{D}} &=& \frac{G_F}{\sqrt{2}} (A \lambda^3) 
	(\rho - i \eta) a_f \nonumber \\ 
B_{\bar{D}} &=& \frac{G_F}{\sqrt{2}} (A \lambda^3) (\rho - i \eta) b_f 
\eea        
where
\bea
a_f &=& - (C_2 + \xi C_1) F_D \left(m_{\Lambda_b} - m_{\Lambda} \right) g_V \nonumber \\
b_f &=& (C_2 + \xi C_1) F_D \left(m_{\Lambda_b} + m_{\Lambda} \right) g_A 
\eea
In Eq. (3), we have also included the pole contribution $b_p$ which is 
given by 
\bea
B_p     &=& \frac{G_F}{\sqrt{2}} (A \lambda^3) b_p \nonumber \\
	&=& - \left[
		\frac{g^{'}\left< \Xi^{'0}_c | H^{pc}_W | \Lambda_b \right>} 
		     {m_{\Lambda_b} - m_{\Xi^{'0}_c}}  
		+ \frac{g \left< \Xi^{0}_c | H^{pc}_W | \Lambda_b \right>} 
		     {m_{\Lambda_b} - m_{\Xi^{0}_c}}  \right]
\eea
where
$g^{'} \equiv g_{\Xi^{'0}_c \Lambda D}$
and 
$g \equiv g_{\Xi^{0}_c \Lambda D}$
are strong coupling constants. In the $W$-exchange model \ct{2}, 
$b_p$ is given by \ct{9} 
\bea
b_p     &=& \left[
		\frac{ - \sqrt{3} g^{'}} {m_{\Lambda_b} - m_{\Xi^{'0}_c}}  
		+ \frac{g} {m_{\Lambda_b} - m_{\Xi^{0}_c}}  \right] d^{'} \nonumber \\
     d^{'} &=& 5 \times 10^{-3} \mbox{GeV}^3
\eea
Note the important fact that 
$\Xi^{'0}_c$ 
and 
$\Xi^{0}_c$ 
poles contribute to the decay
$\Lambda_b \rightarrow \Lambda D^0$ only. This is crucial in discussing 
the CP violation.

Let us symbollically write for the two channels of the decay: 
\be 
\left. |\Lambda_b \right>= \left. R_D | \Lambda D^0 \right> 
			+ \left. R_{\bar{D}} | \Lambda \bar{D}^0 \right>
\ee
where, $R_D = A_D$ or $B_D$ and $R_{\bar{D}} = A_{\bar{D}}$ 
or $B_{\bar{D}}$ for $s$ or $p$-wave respectively. 
From Eq. (8), it is clear that the decay amplitudes for the states 
\bea
	D^0_+ &=& \frac{1}{\sqrt{2}} (D^0 + \bar{D}^0 ), \nonumber \\
	D^0_- &=& \frac{1}{\sqrt{2}} (D^0 - \bar{D}^0 ), 
\eea
which are eigenstates of CP are given by 
\bea 
	A_{\pm} &=& \frac{1}{\sqrt{2}} 
		 \left[ \frac{G_F}{\sqrt{2}} A \lambda^3 \right] 
		 \left[(1 \pm \rho ) \mp i \eta \right] a_f \nonumber \\
	B_{\pm} &=& \frac{1}{\sqrt{2}} 
		 \left[ \frac{G_F}{\sqrt{2}} A \lambda ^3 \right] 
		 \left[ \left( (1 \pm \rho ) \mp i \eta \right) 
			b_f + b_p \right] 
\eea 
It is the presence of the pole contrbution in the interference of the 
amplitudes that make $Im A^{*}_{\pm} B_{\pm}$ non zero, giving rise to 
CP violation. Now $Im A^{*}_{\pm} B_{\pm}$ is proportional to 
$\pm \eta (b_p/b_f)$. If $b_p/b_f$ is not small i.e., if $b_p$ is 
comparable with $b_f$, then $\beta_{\pm}$ may be experimentally 
measurbale. 
 
To fit the experimental branching ratio for the decay
$\Lambda_b \rightarrow \Lambda J/{\psi}$, 
it was shown in reference \ct{1}, that with $C_1$ and $C_2$ as 
given in Eq. (3), the parameter $\xi = 1/N_c$  in 
$C_1 + \xi C_2$ must lie within the 
following limits:
\be
	0 \leq \xi \leq 0.125 \,\, 
	\mbox{or} \,\,  
	0.35 \leq \xi \leq 0.45 
\ee
These limits are consistent with the limit
$0 \leq \xi \leq 0.5$  
suggested by the combined anaylsis of the present CLEO data on 
$B \rightarrow h_1 h_2$ decay \ct{8}. To give a crude estimate for the 
strong coupling constants $g$ and $g^{'}$, $SU(4)$ is used as a 
guide. In $SU(4)$ 
\bea
g^{'} &=& \sqrt{3/2} (g_f - g_d) = \sqrt{3/2} (1 - 2 f) g_{\pi N N} \nonumber \\ 
g &=& - \sqrt{1/2} (g_f + g_d/3) = - \frac{1}{3 \sqrt{2}} (1 + 2 f) g_{\pi N N} \nonumber \\ 
g^{'}/g &=& \sqrt{3} \frac{g_f - g_d}{g _f + g_d/3} 
	= \sqrt{3} \frac{F-D}{F+ D/3}
\eea
Using $F = 0.446, D = 0.815$, we obtain 
\be
g^{'}/g = -0.890, g = - 0.493 g_{\pi \pi N}
\ee
Now $ b_p/{b_f}$ can be easily obtained from Eqs. (5) and (7). 
For $\xi=0, C_2 = - 0.257, F_D = 0.200$ GeV, 
$g_A (m^2_D) = - 0.120, m_{\Lambda_b} = 5.641$GeV, 
$m_{\Lambda} = 1.116$ GeV, 
$m_{\Xi^{'0}_c} = 2.580$ GeV, 
$m_{\Xi^{0}_c} = 2.468$ GeV, 
$d^{'} = 5 \times 10^{-3}$ GeV$^3$,  
$g_{\pi N N} = 13.26$ 
and Eq. (13), we get 
\be
x\equiv b_p/b_f = - 0.64 
\ee
For $\xi = 0.125$, one gets $x = - 1.41$. Due to uncertainty in the  
parameter $\xi$ and in the values of $g^{'}$ and $g$, we treat 
$x$ as a free parameter in the range  
$ - 1 \leq x \leq 1$. 

Using Eqns (4), (5), (7) and (10) and the experimental 
values of the masses given above, we obtain
\be
\frac{\Gamma (\lab \ra \la D^o)}{\Gamma(\lab \ra \la \dbn)}
= \frac{D}{\rho^2 + \eta^2}
\ee
\bea
\alpha_D &=& \frac{-(1+x)}{D} \nonumber\\
\alpha_{\bar{D}} &=& -1\\
D &=&  1+ 0.946 x^2 + 0.473 x^2
\eea
\be
\tan \Delta_{\pm} = \frac{-[r_{\pm} \sin (\de_p - \de_s)
	+x\sin(\de_p -\de_s \pm \hat{\phi},{\hat{\beta}})]}
	{[r_{\pm} \cos (\de_p - \de_s)
	+x\cos(\de_p -\de_s \pm \hat{\phi},{\hat{\beta}})]}
\ee
\be
\tan \phi_{\pm} = \frac{-[r_{\pm} \sin (\de_p - \de_s)
	+x\sin(\de_p -\de_s \pm \hat{\phi},{\hat{\beta}})]}
	{[0.053r_{\pm} 
	-0.940x\cos(\hat{\phi},{\hat{\beta}}) +0.470\frac{x^2}{r_{\pm}}]}
\ee
where
\bea
r_{\pm} &=& \sqrt{(1\pm \rho)^2 + \eta^2}\nonumber\\
\tan \hat{\phi} &=& \frac{\eta}{1+\rho}, \nonumber\\
\tan \hat{\beta} &=&  \frac{\eta}{1-\rho}
\eea
Eqs. (15-20) are main results. Using \ct{1}
\be
\frac{\Gamma (\lab \ra \la \dbn)}{\Gamma(\lab \ra \la J/\Psi)}
= (2.8 \times 10^{-3})(\rho^2 + \eta^2)
\ee
We can express Eq. (15) as

\be
\frac{\Gamma (\lab \ra \la \dbn)}{\Gamma(\lab \ra \la J/\Psi)}
= (2.8 \times 10^{-3})D
\ee
The above branching ratio and the asymmetry parameter $\alpha_D$ for
various values of $x$ are given in Table 1.

From Table 1 one can see  that $\alpha_D$ can be as low as zero 
where as $\alpha_{\bar{D}}=-1$ independent of $x$.

In the absence of final state interaction i.e.
$\de_p-\de_s=0$, the angles $\Delta_{\pm}$ and $\phi_{\pm}$
as function of the parameter $x$ given in Eqs. (18) and (19) are
plotted in Figs. (1) and (2) respectively for the preffered values of
$(\rho,\eta)$ of reference \ct{10} viz $(\rho,\eta)= (0.05,0.36)$.
Of course similar curves ca be obtained for different values of 
$(\rho, \eta)$ allowed by existing data.

Figs. (1) and (2) for the angles $\Delta_{\pm}$ and $\phi_{\pm}$ are
a unique features  of interference of the amplitudees for
$\lab \ra \la D^o$ and $\la \dbn$. The experimental values of $\alpha_D$
would fix the value of the parameter $x$. Once $x$ is known then
Eqs. (18) and (19) [if $\de_p - \de_s) \sim 0$] can be plotted as
function of $(\rho,\eta)$. This may give some intersting information
about $CP$ violation.

However we have estimated
$x=-0.64$ or $-1.41$ for $\xi =0$ or $\xi =0.125$.
For these values of x and $(\rho,\eta)=(0.05,0.36)$, we get
$$x=-0.64,\; \alpha_D=-0.61,\; \Delta_+ = 24^o,\; \Delta_- = -33^o,
\; \phi_+ =26^o,\; \phi_- = -28^o$$ 
$$x=-1.41,\; \alpha_D= 0.68,\; \Delta_+ = -64^o,\; \Delta_- = 59^o,
\; \phi_+ =48^o,\; \phi_- = -53^o$$
 
It is clear that determination of $\alpha_D$ in future experiment will be
by itself interesting even though the experimental observation of 
 $CP$ violation in these decays may be difficult in near future.

I am grateful to Riazuddin for discussions.

\newpage
Table - I 
\vspace{2cm}
\begin{table}
\centering
\begin{tabular}{|l|l|l|} 
\hline
{$x$} & {$\alpha_D$} &     
$\left[\frac{\Gamma (\lab \ra \la \dbn)}{\Gamma(\lab \ra \la J/\Psi)}\right]
\times 10^3$ \\
\hline
-1   &  0     & 1.5 \\ 
-0.5 & - 0.77 & 1.8 \\
0    & -1.0   & 2.8 \\ 
 0.5 & - 0.94 & 4.4 \\ 
1 & - 0.83 & 6.8 \\ \hline
\end{tabular}
\end{table}
{\bf The decay parameter $\alpha_D$ and the braching ratio
$\frac{\Gamma (\lab \ra \la \dbn)}{\Gamma(\lab \ra \la J/\Psi)}$.}
%\begin{description}
%\item{Table I:} The decay parameter $\alpha_D$ and the braching ratio
%$\frac{\Gamma (\lab \ra \la \dbn)}{\Gamma(\lab \ra \la J/\Psi)}$.
%\end{description}

\newpage
\begin{figure}[htb]
\epsfxsize=4 in
\centerline{\epsffile{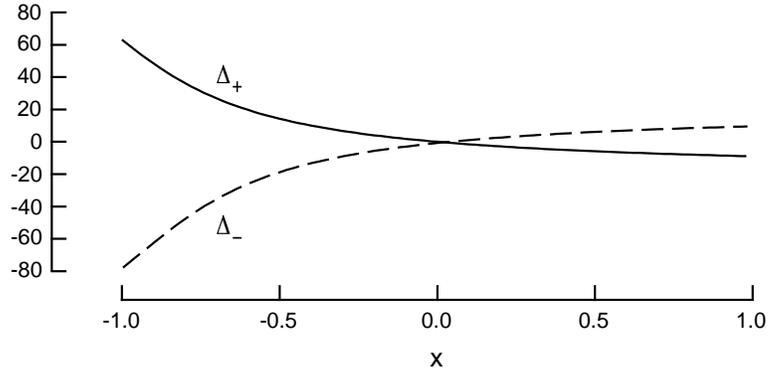}}
\caption{The angle $\Delta_{\pm} = \arctan (-\beta_{\pm}/\alpha_{\pm})$
as a function of $x$ [See Eq. (18), $(\de_p -\de_s)=0$]
for $(\rho, \eta)= (0.05,0.36)$. Solid line gives $\Delta_+$ and dotted 
line gives $\Delta_-$.}
\end{figure}

\begin{figure}[htb]
\epsfxsize=4 in
\centerline{\epsffile{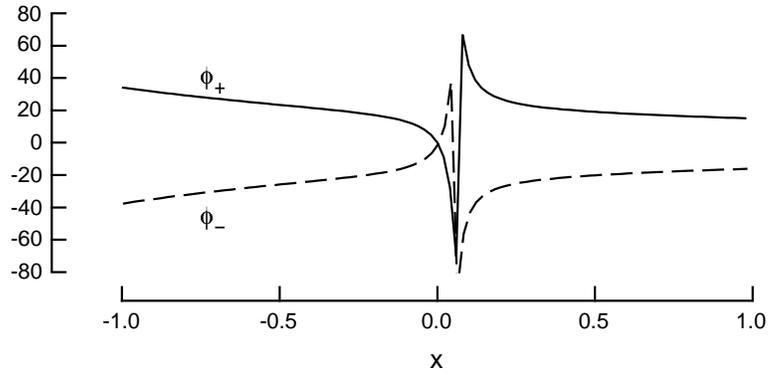}}
\caption{The angle $\phi_{\pm}= \arctan (\beta_{\pm}/\gamma_{\pm})$
as a function of $x$ [See Eq. (19), $(\de_p - \de_s)=0$]
$(\rho,\eta)= (0.05,0.36)$. Solid line gives $\phi_+$ and dotted line 
gives $\phi_-$.}
\end{figure}

%{\bf Fig Captions}\\
%\begin{description}
%\item{Fig1:} The angle $\Delta_{\pm} = \arctan (-\beta_{\pm}/\alpha_{\pm})$
%as a function of $x$ [See Eq. (18), $(\de_p -\de_s)=0$]
%for $(\rho, \eta)= (0.05,0.36)$. Solid line gives $\Delta_+$ and dotted 
%line gives $\Delta_-$.
%
%\item{Fig 2:} The angle $\phi_{\pm}= \arctan (\beta_{\pm}/\gamma_{\pm})$
%as a function of $x$ [See Eq. (19), $(\de_p - \de_s)=0$]
%$(\rho,\eta)= (0.05,0.36)$. Solid line gives $\phi_+$ and dotted line 
%gives $\phi_-$.
%\end{description}

\end{document}